\documentclass[smallextended]{svjour3}

\begin{document}

\title{Ermakov-Lewis invariant in Koopman-von Neumann mechanics}

\author{Abhijit Sen \and Zurab Silagadze}
\institute{Abhijit Sen \at Novosibirsk State University, Novosibirsk 630 090, 
Russia. \email{abhijit913@gmail.com} \and Zurab Silagadze \at Budker Institute 
of Nuclear Physics and Novosibirsk State University, Novosibirsk 630 090, 
Russia. \email{Z.K.Silagadze@inp.nsk.su}}

\date{Received: date / Accepted: date}

\maketitle

\begin{abstract}
In the paper Sci. Rep. 8, 8401 (2018), among other things, the Ermakov-Lewis 
invariant was constructed for the time dependent harmonic oscillator in 
Koopman-von Neumann mechanics. We point out that there is a simpler method 
that allows one to find this invariant.
\end{abstract}

\section{Introduction}
In the interesting paper \cite{1}, the Ermakov-Lewis invariant is used to 
study the time-dependent harmonic oscillator (TDHO) in the framework of 
Koopman-von Neumann (KvN) mechanics \cite{2,3}. To find  the invariant, 
a system of coupled differential equations was obtained, which then
was reduced to a single equation related to the Ermakov equation.
Although this method is quite simple, our goal in this short note is to show 
that there is an even simpler way to find the Ermakov-Lewis invariant for 
TDHO in KvN mechanics.

Recent interest in KvN mechanics is motivated by experiments exploring
the quantum-classical border. To formulate a consistent framework for
a hybrid quantum-classical dynamics is a long-standing problem \cite{3A}.
Its solution, in addition to practical interest, for example, in quantum 
chemistry, can clarify deep conceptual issues in quantum mechanics, such 
as the problem of measurement. An interesting result in this direction was
obtained in \cite{3B}. It was found that the Wigner function in the nonrelativistic
limit turns into the Koopman-von Neuman wave function, which explains why
the Wigner function  is not positive-definite.

Time-dependent harmonic oscillators arise in many quantum mechanical 
systems \cite{3C,3D}. At the same time, the existence of Ermakov-Lewis 
invariants in such systems has attracted much attention \cite{3D}. 
In our opinion, extention of these results to the case of KvN mechanics  
is of considerable interest.

\section{KvN evolution equation for TDHO in new variables}
The KvN evolution equation for TDHO wave-function has the form \cite{1}:
\begin{equation}
i \frac{\partial}{\partial t} \psi(x, p ; t)=\left[\hat{p} \hat{\lambda}_{x}-
k(t) \hat{x} \hat{\lambda}_{p}\right] \psi(x, p ; t),
\label{eq1}
\end{equation}
where $ \hat{\lambda}_{x}$ and $\hat{\lambda}_{p}$ operators satisfy the 
following commutation rules
\begin{equation}
\left[\hat{x},\hat{\lambda}_{x}\right]=\left[\hat{p},\hat{\lambda}_{p}\right]=
i.
\label{eq1A}
\end{equation}
Note that $m=1$ and $\hbar = 1$ was assumed for simplicity.  As Sudarshan
remarked \cite{4}, any KvN-mechanical system can be considered as a hidden 
variable quantum system. Correspondingly,  we will make a slight change in 
notations as follows:
\begin{equation}
x=q , \;\; \lambda _{x}=P,  \;\; \lambda _{p}=-Q,
\label{eq2}
\end{equation}
where $Q$ and $P$ are quantum variables that are hidden for classical observers.
Thus eq.(\ref{eq1}) takes the following form in new notations
\begin{equation}
i \frac{\partial}{\partial t} \psi(q, p ; t)=\left[\hat{p} \hat{P}+k(t) 
\hat{q} \hat{Q}\right] \psi(q, p ; t).
\label{eq3}
\end{equation}
Correspondingly, the KvN Hamiltonian is given by
\begin{equation}
\mathcal{H}=\hat{p} \hat{P}+k(t) \hat{q} \hat{Q}.
\label{eq4}
\end{equation}
Let us make the following canonical transformation
\begin{eqnarray} &&
\hat{q}=\frac{1}{\sqrt{2}}\left(\hat{q}_{1}-\hat{q}_{2}\right), \;\; \hat{Q}=
\frac{1}{\sqrt{2}}\left(\hat{q}_{1}+\hat{q}_{2}\right),\nonumber \\ &&
\hat{p}=\frac{1}{\sqrt{2}}\left(\hat{p}_{1}+\hat{p}_{2}\right), \;\; 
\hat{P}=\frac{1}{\sqrt{2}}\left(\hat{p}_{1}-\hat{p}_{2}\right). 
\label{eq5}
\end{eqnarray}
The transformation is canonical in the sense that it doesn't change the 
canonical form of the commutation relations:
\begin{equation}
\left[\hat{q}_{i}, \hat{q}_{j}\right]=0, \quad\left[\hat{q}_{i}, \hat{p}_{j}
\right]=i \delta_{ij}, \quad\left[\hat{p}_{i}, \hat{p}_{j}\right]=0.
\label{eq6}
\end{equation}
The KvN Hamiltonian when written in new variables 
$\hat{q}_{1},\hat{q}_{2},\hat{p}_{1},\hat{p}_{2}$ 
splits into difference of two Schr\"{o}dinger type quantum Hamiltonians:
\begin{equation}
\mathcal{H} = \left(\frac{\hat{p}_{1}^{2}}{2}+\frac{1}{2} k(t) 
\hat{q}_{1}^{2}\right)-\left(\frac{\hat{p}_{2}^{2}}{2}+\frac{1}{2} k(t)
\hat{q}_{2}^{2}\right) =\mathcal{H}_{1}-\mathcal{H}_{2}.
\label{eq7}
\end{equation}
In the next section we will use this splitting to find the Ermakov-Lewis 
invariant.

\section{Ermakov-Lewis invariant for KvN TDHO}
Let $I$ be the Ermakov-Lewis invariant for KvN TDHO. It must satisfy the 
following equation
\begin{equation}
\frac{d \hat{I}}{d t}=\frac{\partial \hat{I}}{\partial t}-i 
[\hat{I}, \mathcal{H}]=0.
\label{eq8}
\end{equation} 
Since $\mathcal{H}=\mathcal{H}_{1}(\hat{q}_{1},\hat{p}_{1})- 
\mathcal{H}_{2} (\hat{q}_{2},\hat{p}_{2})$, $\hat{I}$ will have the form 
$\hat{I}=\hat{I}_{1}+\hat{I}_{2}$, where $\hat{I}_{1}$ and $\hat{I}_{2}$ are
the usual quantum mechanical Ermakov-Lewis invariants associated respectively
with $\mathcal{H}_{1}(\hat{q}_{1},\hat{p}_{1})$ and $\mathcal{H}_{2} (\hat{q}_{2},
\hat{p}_{2})$. However, there is a subtlety associated with  the minus sign in
$\mathcal{H}=\mathcal{H}_{1}(\hat{q}_{1},\hat{p}_{1})- 
\mathcal{H}_{2} (\hat{q}_{2},\hat{p}_{2})$, which indicates that in the second
Ermakov-Lewis invariant we should assume time-reversal. More formally, we have
\begin{equation}
\frac{\partial \hat{I}}{\partial t}-i[\hat{I}, \mathcal{H}] =
\left(\frac{\partial I_{1}}{\partial t}-i\left[I_{1}, 
\mathcal{H}_{1}\right]\right)+\left(\frac{\partial I_{2}}{\partial t}+i
\left[I_{2}, \mathcal{H}_{2}\right]\right)=0. 
\label{eq9}
\end{equation}
The individual terms in the brackets must vanish. However for
\begin{equation}
\frac{\partial I_{1}}{\partial t}-i\left[I_{1}, \mathcal{H}_{1}\right]=0,
\label{eq10}
\end{equation}
the corresponding quantum-mechanical invariant is well known \cite{5}
\begin{equation}
I_{1}=\frac{1}{2}\left[\left(\frac{\hat{q}_{1}}{\rho}\right)^{2}+
\left(\hat{p}_{1} \rho-\dot{\rho} \hat{q}_{1}\right)^{2}\right],
\label{eq11}
\end{equation}
where $\rho$ obeys the Ermakov equation \cite{6}
\begin{equation}
\ddot\rho+k(t)\rho=\rho^{-3}.
\label{eq11A}
\end{equation}
As for the equation containing  $\hat{I}_{2}$, define $\tau=-t$ and the equation
takes the form:
\begin{equation}
\frac{\partial I_{2}}{\partial \tau}-i \left[I_{2}, 
\mathcal{H}_{2}\right]=0.
\label{eq12}
\end{equation}
It is clear that the corresponding invariant is 
\begin{equation}
I_{2}=\frac{1}{2}\left[\left(\frac{\hat{q}_{2}}{\rho}\right)^{2}+
\left(\hat{p}_{2} \rho-\rho^{\prime} \hat{q}_{2}\right)^{2}\right],
\label{eq13}
\end{equation}
where $\rho'=\frac {\partial \rho }{\partial \tau }$. Restoring the derivatives
with respect to $t$, we get
\begin{equation}
I_{2}=\frac{1}{2}\left[\left(\frac{\hat{q}_{2}}{\rho}\right)^{2}+
\left(\hat{p}_{2} \rho+\dot{\rho} \hat{q}_{2}\right)^{2}\right].
\label{eq14}
\end{equation}
Thus the Ermakov-Lewis invariant $I$ in KvN mechanics takes the following 
form, after using eq.(\ref{eq11}) and eq.(\ref{eq14}) and re-writing the
result in terms of  $\hat{q},\hat{Q},\hat{p}$ and $\hat{P}$:
\begin{equation}
I=\frac{\hat{Q}^{2}}{2 \rho^{2}}+\frac{\hat{q}^{2}}{2 \rho^{2}}+\frac{1}{2}
\left\{(\dot{\rho} \hat{q}-\rho \hat{p})^{2}+(\dot{\rho} \hat{Q}- \rho 
\hat{P})^{2}\right\}.
\label{eq15}
\end{equation}
Using the correspondence given by eq.(\ref{eq2}), and re-arranging the above 
expression, we get the invariant in the form found in \cite{1}:
\begin{equation}
\hat{I}=\frac{1}{2}\left[\frac{\hat{x}^{2}}{\rho^{2}}+(\dot{\rho} \hat{x}-
\rho \hat{p})^{2}+\frac{\hat{\lambda}_{p}^{2}}{\rho^{2}}+\left(\dot{\rho} 
\hat{\lambda}_{p}+\rho \hat{\lambda}_{x}\right)^{2}\right].
\label{eq16}
\end{equation}

\section{Conclusions}
We have  shown that one can find the Ermakov-Lewis invariant
in the case of KvN mechanics using the well-known quantum-mechanical 
expression for this invariant \cite{5} and some simple algebra. 
In this method, there is no need to consider any coupled differential equations.
However, it is less general than the method  considered in \cite{1}, which, 
in principle, can be applied to any potential, even if it does not allow 
$\mathcal{H}$ to split in a way described in this note.

\end{document}